\documentclass[aps,prb,superbib,superscriptaddress,amsfonts,preprint]{revtex4}

\usepackage{graphicx}
\usepackage{epsfig,tabularx}
\usepackage{amssymb,amsmath}
\usepackage{multirow,xcolor}
\usepackage{hyperref}
\usepackage[papersize={8.5in,11in}]{geometry}
\hypersetup{
    colorlinks = true,
    linkcolor = blue,
    linkbordercolor = white,
    citecolor=blue,
}
\usepackage{subfig,caption}
\newcommand{\ket}[1]{\ensuremath{\left|#1\right\rangle}}

\newcommand{\bracket}[2]{\ensuremath{\left\langle #1 \middle| #2 \right\rangle}}
\newcommand{\matrixel}[3]{\ensuremath{\left\langle #1 \middle| #2 \middle| #3 \right\rangle}}

\newcommand{\bm}[1]{\boldsymbol{#1}}

\newcommand{\refeq}[1]{Eq. \ref{#1}}
\newcommand{\refsec}[1]{Sec. \ref{#1}}
\newcommand{\refapdx}[1]{Appendix \ref{#1}}
\newcommand{\reffig}[1]{Fig. \ref{#1}}

\newcommand{\refeqs}[1]{Eqs. \ref{#1}}

\newcommand{\reffigs}[1]{Figs. \ref{#1}}

\begin{document}
\title{ An extension of the fewest switches surface hopping algorithm to complex Hamiltonians and photophysics in magnetic fields: Berry's phase and ``magnetic" forces }

\author{Gaohan Miao}
\affiliation{Department of Chemistry, University of Pennsylvania, Philadelphia, Pennsylvania 19104, USA}
\author{Nicole Bellonzi}
\affiliation{Department of Chemistry, University of Pennsylvania, Philadelphia, Pennsylvania 19104, USA}
\author{Joseph Subotnik}
\email{subotnik@sas.upenn.edu}
\affiliation{Department of Chemistry, University of Pennsylvania, Philadelphia, Pennsylvania 19104, USA}

\begin{abstract}
    We present a preliminary extension of the fewest switches surface hopping (FSSH) algorithm to the case of complex Hamiltonians as appropriate for modeling the dynamics of photoexcited molecules in magnetic fields. 
    We make {\em ans\"{a}tze} for the direction of momentum rescaling and we account for Berry's phase effects through ``magnetic'' forces as applicable in the adiabatic limit. 
    Because Berry's phase is a nonlocal, topological characteristic of a set of entangled potential energy surfaces, we find that Tully's {\em local} FSSH algorithm can only partially capture the correct physics.
\end{abstract}

\maketitle

\section{Introduction} \label{sec: introduction}
	Fewest switches surface hopping (FSSH)\cite{tully_molecular_1990}
	has been a very powerful tool for simulating non-adiabatic dynamics over the last thirty years.\cite{oberhofer_charge_2017,habenicht_nonradiative_2008,nelson_nonadiabatic_2011,sterpone_nonadiabatic_2009,nelson_nonadiabatic_2014} 
	The basic idea of the FSSH algorithm is to run stochastic dynamics on electronic adiabats, with stochastic switches between adiabats to account for electronic relaxation; in the spirit of Pechukas's force,\cite{tully_nonadiabatic_1991,pechukas_time-dependent_1969} one rescales momenta in the direction of the derivative coupling whenever a hop between surfaces occurs.
    The algorithm has been shown to successfully capture both the short time dynamics of non-adiabatic systems\cite{landry_quantifying_2014,jain_surface_2015} as well as (their) long time equilibrium properties.\cite{parandekar_mixed_2005}
    At the same time, the cost of FSSH is quite modest\cite{tully_molecular_1990,jain_efficient_2016,barbatti_--fly_2007}.
Of course, Tully's algorithm has a few well-known shortcomings: 
$(i)$ the original algorithm did not treat wave packet separation correctly, and thus did not model decoherence;\cite{schwartz_quantum_1996,prezhdo_evaluation_1997,wong_dissipative_2002,wong_solvent-induced_2002,fang_comparison_1999,fang_improvement_1999,hack_electronically_2001,volobuev_continuous_2000,jasper_electronic_2005,prezhdo_mean-field_1997}
$(ii)$  the algorithm does not treat recoherence correctly;\cite{subotnik_understanding_2016}
$(iii)$ the algorithm does not include
any nuclear quantum effects.\cite{craig_quantum_2004,jang_derivation_1999}  Of the problems above, item $(i)$ has been discussed extensively in the literature and can largely be corrected; items $(ii)$ and $(iii)$ are largely intractable with classical, {\em non-interacting} trajectories.\cite{donoso_quantum_2001,donoso_simulation_2003} Nevertheless,
as a testament of the algorithm's value, FSSH is routinely applied today to simulate non-adiabatic dynamics including photochemical processes\cite{muller_surface-hopping_1997,landry_quantifying_2014,nelson_nonadiabatic_2011}, scattering\cite{shenvi_dynamical_2009,golibrzuch_importance_2014,jain_efficient_2016}, and charge transfer in solution\cite{fang_comparison_1999,schwerdtfeger_nonadiabatic_2014}. 

Interestingly, of all of the applications listed above, there is one glaring omission. To our knowledge,
no one has yet used FSSH to study non-adiabatic dynamics for molecular systems with spin 
degrees of freedom in strong magnetic fields. More generally, to our knowledge, no one has yet 
extended the FSSH algorithm to treat complex (rather than real-valued) electronic Hamiltonians.
When one considers such an extension, several obvious questions arise, including: $(a)$ how should one incorporate
geometric phases semiclassically,\cite{berry_quantal_1984,baer_beyond_2006,yarkony_diabolical_1996,yarkony_current_1996} given that geometric phase is a nonlocal, topological property? $(b)$ how should one choose the direction of momentum rescaling when the derivative coupling is complex, and there is no unique real vector to isolate? The answers are not obvious.
\par
With this background in mind, the goal of this paper is to propose one possible set of answers
and a possible extension of FSSH to the case of complex Hamiltonians. We will find that
our current implementation of FSSH behaves reasonably well, though one clearly loses some accuracy when moving
from the case of real to complex Hamiltonians. In particular, because of topological phase effects, we will show
that obvious limitations arise for any algorithm (like FSSH) based on independent, spatially local and time local
 trajectories.
	This paper is structured as follows:
    In \refsec{sec: methods}, we introduce our several {\em ans\"{a}tze} for the FSSH algorithm in 
the presence of a  complex Hamiltonian.
	In \refsec{sec: simulation details}, we make clear our simulation details.
    In \refsec{sec: results}, we present our results.
    In \refsec{sec: discussion_and_conclusion}, we interpret our numerical results and give a simple explanation for how geometric phase effects appear in surface hopping, and we propose a general extension of the FSSH algorithm.
    Finally, in \refsec{sec: summary} and \refsec{sec: open questions}, we summarize the paper and present some open questions, respectively.
	As far as notation is concerned, below we use
	bold characters (e.g. $\bm{r}$) to denote  vectors, and we use 
	plain characters (e.g. $H$) to denote either scalars or operators.

\section{Methods} \label{sec: methods}

\subsection{Real Hamiltonians} \label{sec: real hamiltonian}
        Let us now briefly review the FSSH algorithm.  As originally conceived, the FSSH approach is applicable to the case of real electronic Hamiltonians. Without loss of generality, consider a real two-by-two Hamiltonian ({\em i.e.} a Hamiltonian with two electronic states) of the form

\begin{equation}\label{eq: real Hamiltonian}
\begin{aligned}
H(\bm{r}) = \left( 
\begin{array}{cc}
V_{00}(\bm{r}) & V_{01}(\bm{r}) \\
V_{10}(\bm{r}) & V_{11}(\bm{r})
\end{array}
\right)
\end{aligned}
\end{equation}

Here, $\bm{r}$ is a nuclear coordinate.  To simulate semiclassical dynamics with quantum electronic states and classical nuclei, according to FSSH, one first diagonalizes the electronic Hamiltonian and computes
adiabatic energies $E_0(\bm{r}), E_1(\bm{r})$, forces $\bm{F}_0(\bm{r}), \bm{F}_1(\bm{r})$, and derivative couplings $\bm{d}_{01}(\bm{r})$. Thereafter, one runs an ensemble of independent trajectories, initialized so as to correspond the correct Wigner distribution at time zero\cite{wigner_quantum_1932,subotnik_understanding_2016,landry_communication:_2013,subotnik_can_2013,kapral_surface_2016}. Each trajectory is evolved along a single adiabatic surface, with  equations of motion:

\begin{equation}\label{eq: FSSH propagation}
\begin{aligned}
    \dot{\bm{r}} &= \frac{\bm{p}}{m}\\
    \dot{\bm{p}} &= \bm{F}_j
\end{aligned}
\end{equation}

Here $j$ is the active surface for a given trajectory. Occasionally, trajectories switch from one surface to the other. 
For example, Tully proposed\cite{tully_molecular_1990} that a trajectory on surface $0$ switches to surface $1$ with rate
\begin{eqnarray} \label{eq: FSSH hop rate}
g_{0\rightarrow 1} &= \max\left[0, \Delta t \frac{\dot{\rho}_{11}}{\rho_{00}} \right]
\end{eqnarray}
Here $\rho_{jk} \equiv c_jc_k^*$ are density matrix elements, and $(c_0, c_1)$  is the electronic wavefunction.  Whenever a particle switches surfaces, in order to conserve energy, one rescales the momentum in the direction of the derivative coupling, $\bm{d}_{01}(\bm{r})$.\cite{herman_generalization_1982} There are many existing references in the literature where one can learn more details of the FSSH algorithm\cite{tully_molecular_1990,fang_comparison_1999,barbatti_nonadiabatic_2011,subotnik_understanding_2016}, beginning with Tully's original paper. \cite{tully_molecular_1990}

\subsection{Complex Hamiltonians} \label{sec: complex hamiltonian}
At this point, we come to the heart of the matter. Consider a situation whereby a particle with spin interacts with a magnetic field and there are two possible electronic states.
Because of the magnetic field, the electronic Hamiltonian will no longer be real-valued.\cite{mead_noncrossing_1979,yabushita_spinorbit_1999,domcke_conical_2004, belcher_gradients_2011}
Instead, the electronic Hamiltonian will be complex and therefore violate time reversibility. In other words, in \refeq{eq: real Hamiltonian}, $V_{01}(\bm{r})$ can have both real and imaginary parts,\cite{domcke_conical_2004} and $V_{10}(\bm{r}) = V_{01}^*(\bm{r})$.   
For this situation, FSSH is not well defined and two obvious problems present themselves.

\begin{enumerate}
	\item  First, note that FSSH depends critically on the existence of adiabatic states.
	Now, it is well known that, in the presence of conical intersections, adiabatic electronic states cannot be globally defined, even for real electronic Hamiltonians.\cite{domcke_conical_2004} Nevertheless, even though FSSH does not account
	for geometric phase, the algorithm is largely able to model dynamics through conical intersections, as has been documented in detail previously\cite{ryabinkin_geometric_2013,ryabinkin_when_2014,ryabinkin_geometric_2017,gherib_why_2015}. That being said, for the present case of a complex Hamiltonian, one must always worry: How should one best choose the 
	sign of the wavefunctions, when the sign has a true complex phase and not just a plus/minus? And how should one best incorporate Berry's phase effects\cite{berry_quantal_1984,shankar_principles_2012} semiclassically?
	
	\item The second obvious question is: What is the (real-valued) direction for rescaling momentum? Obviously
	$Re(\bm{d}_{01})$ is not acceptable as this quantity depends on the choice of phase for the adiabatic electronic states. 
    Furthermore, for a practical FSSH calculation, we must be able to compute this direction using only local information at a single nuclear geometry.
\end{enumerate}
	With these two questions in mind, we will propose a few simple and robust extensions of FSSH to complex Hamiltonians. 

    \subsubsection{``Magnetic Force" {\em Ansatz}} \label{sec: magnetic force}
    As far as the changing (Berry) phase of the adiabatic electronic states, it is well known that the  Berry curvature near the crossing region can be transformed into an effective magnetic field that is applicable in the adiabatic limit.\cite{shankar_principles_2012,sakurai_modern_2014}
    Thus, to incorporate Berry's phase effects into FSSH dynamics, we propose that, when a trajectory is moving on adiabatic surface $j$ near a crossing point, we will allow each FSSH trajectory to feel this extra ``magnetic force'':
		\begin{equation}\label{eq: magnetic force}
		\begin{aligned}
			\bm{F}^{mag}_{j} &= \hbar\frac{\bm{p}}{m} \times \bm{B}_j
		\end{aligned}
		\end{equation}
	Here,  $\bm{B}_j$ is defined to be the Berry curvature\cite{berry_quantal_1984,mead_determination_1979}
		\begin{equation}\label{eq: berry curvature}
		\begin{aligned}
			\bm{B}_j &= \bm{\nabla} \times \left(i\matrixel{\psi_j}{\bm{\nabla}}{\psi_j}\right) = -i\sum_{k\neq j} \bm{d}_{jk} \times \bm{d}_{kj}
		\end{aligned}
		\end{equation}
	Substituting \refeq{eq: berry curvature} into \refeq{eq: magnetic force}, and utilizing the identity $\bm{d}_{jk} = -\bm{d}_{kj}^*$, we find
		\begin{equation}\label{eq: magnetic force expr2}
		\begin{aligned}
			\bm{F}^{mag}_{j} &= 2\hbar\mbox{Im}{\sum_{k \neq j}\left[\bm{d}_{jk}(\frac{\bm{p}}{m}\cdot\bm{d}_{kj})\right]}
		\end{aligned}
		\end{equation}

    In the end, for an FSSH simulation moving along adiabat $j$, we will assume that the ``magnetic" force $\bm{F}^{mag}_j$ should simply be added to the total adiabatic, Born-Oppenheimer force in \refeq{eq: FSSH propagation}. 
    Note that $\bm{p} \cdot \bm{F}^{mag}_j = 0$, so that this extra ``magnetic" force does not break energy conservation, but rather turns the direction of momentum. 
    Note further that this  ``magnetic'' force disappears for the case of a real-valued Hamiltonian, where the derivative coupling $\bm{d}_{jk}$ is real.
    Interestingly, for a two state problem, \refeq{eq: magnetic force expr2} implies that $\bm{F}^{mag}_0 = -\bm{F}^{mag}_1$.

	\subsubsection{Direction of Momentum Rescaling} \label{sec: direction of momentum rescaling}
    In order to extend FSSH to the case of a complex Hamiltonian, we must find an appropriate direction for momentum rescaling, $\bm{n}_{jk}$, when a hop between adiabats $j \rightarrow k$ occurs. 
    To be appropriate, this direction vector must satisfy at least three constraints: 
    $(i)$ $\bm{n}_{jk}$ must be real; 
    $(ii)$ $\bm{n}_{jk}$ should not depend on the phase of the derivative coupling $\bm{d}_{jk}$; 
    $(iii)$ $\bm{n}_{jk}$ must reduce to $\bm{d}_{jk}$ when the complex part of the Hamiltonian is removed.
    Furthermore, we must be able to construct this direction with only local information at a single nuclear geometry; we cannot assume that we have any information about a global reaction coordinate.

	With these constraints in mind, the following three {\em ans\"{a}tze} for $\bm{n}_{jk}$ are possibilities:
	\begin{itemize}
		\item Method \#1: ``$Re(\bm{d}(\bm{v}\cdot\bm{d}))$" 
		\par
		Because the magnetic force is independent of phase, the following {\em ansatz} would appear reasonable:
		\begin{equation}\label{eq: rescaling direction real dvd ansatz}
		\begin{aligned}
			\bm{n}_{jk} &= \mbox{Re} \left[\bm{d}_{jk} \left(\frac{\bm{p}}{m} \cdot \bm{d}_{kj} \right)\right]
		\end{aligned}
		\end{equation}
        Note the strong connection between the magnetic force (\refeq{eq: magnetic force expr2}) and $\bm{n}_{jk}$ here: 
        According to \refeq{eq: rescaling direction real dvd ansatz}, the real part of $\left[\bm{d}_{jk} \left(\frac{\bm{p}}{m} \cdot \bm{d}_{kj} \right)\right]$ would act as a direction for momentum rescaling while the imaginary part acts as a magnetic force that modifies motion along a given adiabat (see \refeq{eq: magnetic force expr2}). 
		
        \item Method \#2: ``$Re(e^{i\eta} \bm{d})$"
		\par
        Another option for the rescaling direction $\bm{n}_{jk}$ is the real part of the derivative coupling with a robust phase factor. To this end, one can choose
		\begin{equation}\label{eq: rescaling direction real deieta ansatz}
		\begin{aligned}
            \bm{n}_{jk} &= Re(e^{i\eta} \bm{d}_{jk}) ,
		\end{aligned}
		\end{equation}
            where for every coordinate $\bm{r}$, $\eta$ is chosen so as to maximize the vector norm $||Re(e^{i\eta} \bm{d}_{jk})||^2$.
            Note that, unlike Method \#1, this {\em ansatz} for $\eta$ does not depend on any dynamical properties of a given trajectory.
            
        \item Method \#3: ``Average $\bm{d}$"
        \par
            One last possibility is the averaged derivative coupling (divided by $2i$) \footnote{Likely, $\bm{n}_{jk} = Re(\rho_{jk}\bm{d}_{kj})$ is another possibility.}
		\begin{equation}\label{eq: rescaling direction real deieta ansatz}
		\begin{aligned}
            \bm{n}_{jk} &= \frac{1}{2i}(\rho_{jk}\bm{d}_{kj} + \rho_{kj}\bm{d}_{jk}) = Im(\rho_{jk}\bm{d}_{kj}) 
		\end{aligned}
		\end{equation}
        Like Method \#1, this {\em ansatz} depends on the dynamics of a given trajectory. 
        However, whereas Method \#1 makes use of the nuclear momentum, Method \#3 makes use of the electronic density matrix to construct the rescaling direction.
	\end{itemize}
    In practice, as shown in the \refapdx{apdx: method 3 results}, Method \#3 performs very poorly,
    \footnote{Similarly, $\bm{n}_{jk} = Re(\rho_{jk}\bm{d}_{kj})$ does not perform well.}
    and so below we will focus exclusively on Methods \#1 and \#2.

    Throughout this paper, there is one nuance worth reporting. 
    When running FSSH calculations, one needs to choose appropriate phases for eigenvectors. 
    To choose these phases, one can use either
    $(i)$ eigenvectors computed on the fly, whereby the phase of a given set of eigenvectors are aligned with the eigenvectors at previous time step by ``parallel transport" ({\em i.e.} $\bracket{\psi_i(t)}{\psi_i(t+dt)} \approx 1$); or
    $(ii)$ analytical eigenvectors (see below in \refeq{eq: eva evt dc anal}) for which a global phase is assigned (whenever possible).
    In our FSSH calculations, we find that as long as we initialize the system in a consistent fashion, we can use either phase convention, the difference between $(i)$ and $(ii)$ is negligible. 
    For results below, all FSSH data are implemented using option $(i)$.

\section{Simulation Details} \label{sec: simulation details}
\begin{figure}[h!]
    	\centering
    	\includegraphics[width=4in]{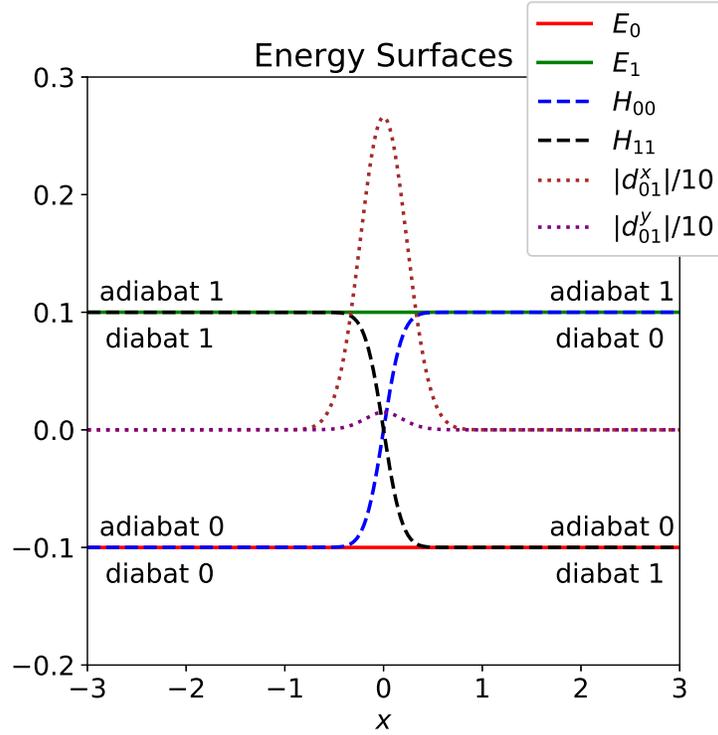}
    	\caption{Surfaces for the Hamiltonian in \refeq{eq: Hamiltonian}. Parameters are: $A = 0.1$, $B = 3.0$, $W = 0.3$. Solid lines are the adiabatic surfaces, which are flat; dashed lines are the diabatic surfaces; dotted lines are the absolute values of the derivative coupling along each direction. }	
    	\label{fig: surf}
    \end{figure}

    Consider a simple 2-D system with the following general Hamiltonian:
        \begin{equation}\label{eq: Hamiltonian}
            \begin{aligned}
                H &= A
                \begin{bmatrix}
                    -\cos{\theta(x,y)} & \sin{\theta}e^{i\phi(x,y)} \\
                    \sin{\theta(x,y)}e^{-i\phi(x,y)} & \cos{\theta(x,y)} \\
                \end{bmatrix} \\
            \end{aligned}
        \end{equation}
    For a simple model, we define the functions $\theta(x,y)$ and $\phi(x,y)$ to be:
        \begin{equation}\label{eq: Theta_phi_def}
            \begin{aligned}
                \theta &\equiv \frac{\pi}{2}\left(erf(Bx) + 1\right) \\
                \phi &\equiv Wy
            \end{aligned}
        \end{equation}
    Here $A$, $B$, $W$ are constants.  In \reffig{fig: surf}, we plot the diabats, adiabats and derivative couplings. Note that the adiabats are completely flat which will make all FSSH results easier to interpret. 
    For this Hamiltonian, one may solve for the eigenvalues, eigenvectors, and the derivative couplings analytically:
    \begin{equation}\label{eq: eva evt dc anal}
    \begin{aligned}
    \lambda_0 &= -A \\
    \lambda_1 &= A \\
    \psi_{0} &= 
    \begin{bmatrix}
    \cos{\frac{\theta}{2}}e^{i\phi} \\
    -\sin{\frac{\theta}{2}}
    \end{bmatrix}
    \\
    \psi_{1} &=
    \begin{bmatrix}
    \sin{\frac{\theta}{2}}e^{i\phi} \\
    \cos{\frac{\theta}{2}}
    \end{bmatrix}
    \\
    \bm{d}_{01} &= \frac{\bm{\nabla}\theta}{2} + i\frac{\bm{\nabla}\phi}{2}\sin{\theta} = \left(\frac{\partial_x\theta}{2}, \frac{i\sin{\theta}\partial_y\phi}{2}\right)
    \end{aligned}
    \end{equation}
    Note that, with the choice of adiabats in \refeq{eq: eva evt dc anal},  $\bm{d}_{01}$ is composed of two components: a real component in the direction of the crossing  ($\bm{\nabla}\theta$) and an imaginary component in the direction of the gradient of the phase of the diabatic coupling ($\bm{\nabla}\phi$). 
    For all dynamics reported below, we initialize Gaussian wave packets on the upper surface
	    \begin{equation}\label{eq: initial exact}
	    \begin{aligned}
	    	\Psi_0(\bm{r})（&= 0 \\
            \Psi_1(\bm{r})（&= e^{\frac{i}{\hbar}\bm{r}\cdot\bm{p}_{init}}e^{-\frac{|\bm{r}-\bm{r}_{init}|^2}{\sigma^2}}
	    \end{aligned}
	    \end{equation}
    Here $\bm{p}_{init}$ and $\bm{r}_{init}$ are the initial momentum and position, respectively; $\sigma$ is the spread of the initial wave packet over real space. 
    For exact quantum calculations, the wave packets are propagated with the Schr\"{o}dinger equation using the fast Fourier transform technique\cite{kosloff_fourier_1983}. 
    For the surface hopping algorithm, $10^7$ trajectories are sampled from the Wigner distribution corresponding to \refeq{eq: initial exact}. Each semiclassical trajectory is propagated according to the (modified) FSSH algorithm with an {\em ansatz} for the rescaling direction as described above. 
    For a particle moving in the 2-D plane, the magnetic forces are of the following form:
    \begin{equation} \label{eq: Fmag anal}
    \begin{aligned}
    \bm{F}^{mag}_{1} &= 2\hbar\mbox{Im}{\left[\bm{d}_{10}(\frac{\bm{p}}{m}\cdot\bm{d}_{01})\right]} = \frac{\hbar}{2m}\partial_x\theta\partial_y\phi\sin{\theta}\left(-p^y, p^x\right) \\
    \bm{F}^{mag}_{0} &= \frac{\hbar}{2m}\partial_x\theta\partial_y\phi\sin{\theta}\left(p^y, -p^x\right)
    \end{aligned}
    \end{equation}
    For Method \#1, the rescaling direction is  
    	\begin{equation}\label{eq: method 1 in model}
    	\begin{aligned}
            \bm{n}_{01} = \left( (\partial_x \theta)^2 p^x, (\partial_y \phi \sin{\theta})^2 p^y \right)
    	\end{aligned}
    	\end{equation}
    For Method \#2, we would ideally like to choose the direction $\bm{\nabla}\theta$, {\em i.e.} the $x$-direction, which we presume is the classical reaction coordinate. Unfortunately, with an arbitrary phase possible when delineating eigenstates, and without the knowledge of a global potential energy surface, isolating $\bm{\nabla}\theta$ is non-trivial.
    In the present case (for a general $\bm{d}$, see \refapdx{apdx: method 2 with a general derivative coupling}), the vector norm $f(\eta) = ||Re(e^{i\eta} \bm{d}_{01})||^2$ becomes
    	\begin{equation}\label{eq: Re d eieta}
    	\begin{aligned}
            f(\eta) &= \frac{1}{2}||Re\left(e^{i\eta} \left(\bm{\nabla}\theta + i\bm{\nabla}\phi\sin{\theta}\right)\right)||^2
    	\end{aligned}
    	\end{equation}
    Maximizing the above expression using $\bm{\nabla}\theta \cdot \bm{\nabla}\phi = 0$, Method \#2 chooses the rescaling direction to be:
    	\begin{equation}\label{eq: method 2 in model}
    	\begin{aligned}
    	\bm{n}_{01} = \left\{
    	\begin{array}{ll}
    	(1,0) & \mbox{ when } (\partial_x\theta)^2 > (\sin{\theta}\partial_y\phi)^2 \\
    	(0,1) & \mbox{ when } (\partial_x\theta)^2 < (\sin{\theta}\partial_y\phi)^2\\
    	\end{array} 
    	\right. 
    	\end{aligned}
    	\end{equation}
    For most parameters below (except \reffig{fig: largeW_ang0_trans}), we will usually operate in the regime whereby $(\partial_x\theta)^2 > (\sin{\theta}\partial_y\phi)^2$, and so $\bm{n}_{01}$ will be in the $x$-direction. 
    For our other parameters, we choose $B = 3.0$, $\bm{r}_{init} = (-3, 0)$, $\sigma = 1.0$. 

    \section{Results} \label{sec: results}
    We begin by investigating scattering processes where the average incoming momentum is along the $x$-direction: $\bm{p}_{init} = (p^x_{init}, 0)$.
    Because we initialize all dynamics to begin in the $x$-direction, we can learn about Berry's phase effects by monitoring all dynamics in the $y$-direction.
	\begin{figure}[h!]
		\centering
		\includegraphics[width=6in]{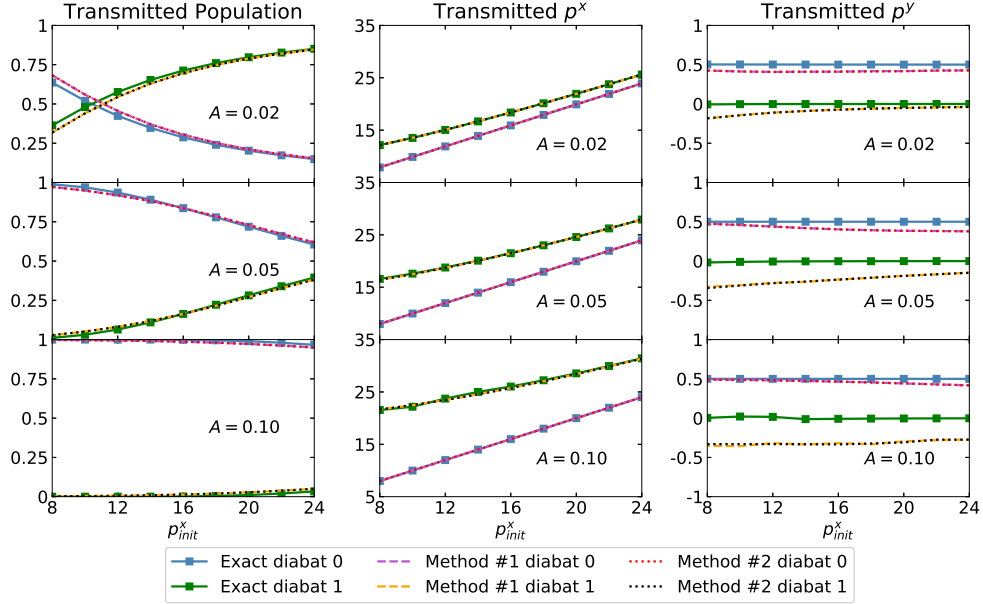}
        \caption{
            Scattering population and momenta with $W = 0.5$. 
            Left: Transmitted population distribution on the diabatic surfaces after scattering as a function of initial momentum along the $x$-direction, $p^x_{init}$.
            Middle: The $x$-direction momentum on the diabatic surfaces as a function of $p^x_{init}$.
            Right: The $y$-direction momentum on the diabatic surfaces as a function of  $p^x_{init}$. 
            The system is initialized with $\bm{p}_{init} = (p^x_{init}, 0)$. 
            FSSH predicts the correct population as well as the $x$-direction momentum on each diabatic surface, while FSSH is only partially correct for the $y$-direction momentum. 
            The difference between two rescaling {\em ans\"{a}tze} is negligible here. }
		\label{fig: smallW_ang0_trans}
	\end{figure}
	In \reffig{fig: smallW_ang0_trans}, we plot the final population and average momentum along the $x$-direction and the $y$-direction for each diabat as a function of initial $x$ momentum, $p^x_{init}$. 
    As far as electronic populations are concerned, the case $A = 0.02$ would appear to be in the diabatic regime at large velocities, where a significant percentage of trajectories stay on the initial diabat (diabat $1 \rightarrow 1$); the case $A = 0.1$ would appear to be in the adiabatic regime, where most trajectories stay on the initial adiabat (diabat $1 \rightarrow 0$).
	The exact quantum dynamics results give a simple interpretation of Berry's phase effect: motion on a given diabat ({\em i.e.} a switch of adiabat) does not lead to a finite momentum in the $y$-direction. 
    By contrast, motion on two different diabats (no switch of adiabats) {\em does} lead to a finite momentum. All momentum changes are identical (with a value of $W = 0.5$), irrespective of the values of $A$ and $p^x_{init}$. 
	\par
	Let us now turn to FSSH.
    For the population distribution and the $x$-direction momentum, we find that the modified surface hopping algorithm does capture the correct results (for both Methods \#1 and \#2). 
    However, for the $y$-direction momentum, while the FSSH result is good for the case of small $A$, its error increases when $A$ is tuned to larger values.
    Again, there is no significant difference between the two rescaling {\em ans\"{a}tze}.
	\par
\begin{figure}[h!]
    	\centering
    	\includegraphics[width=6in]{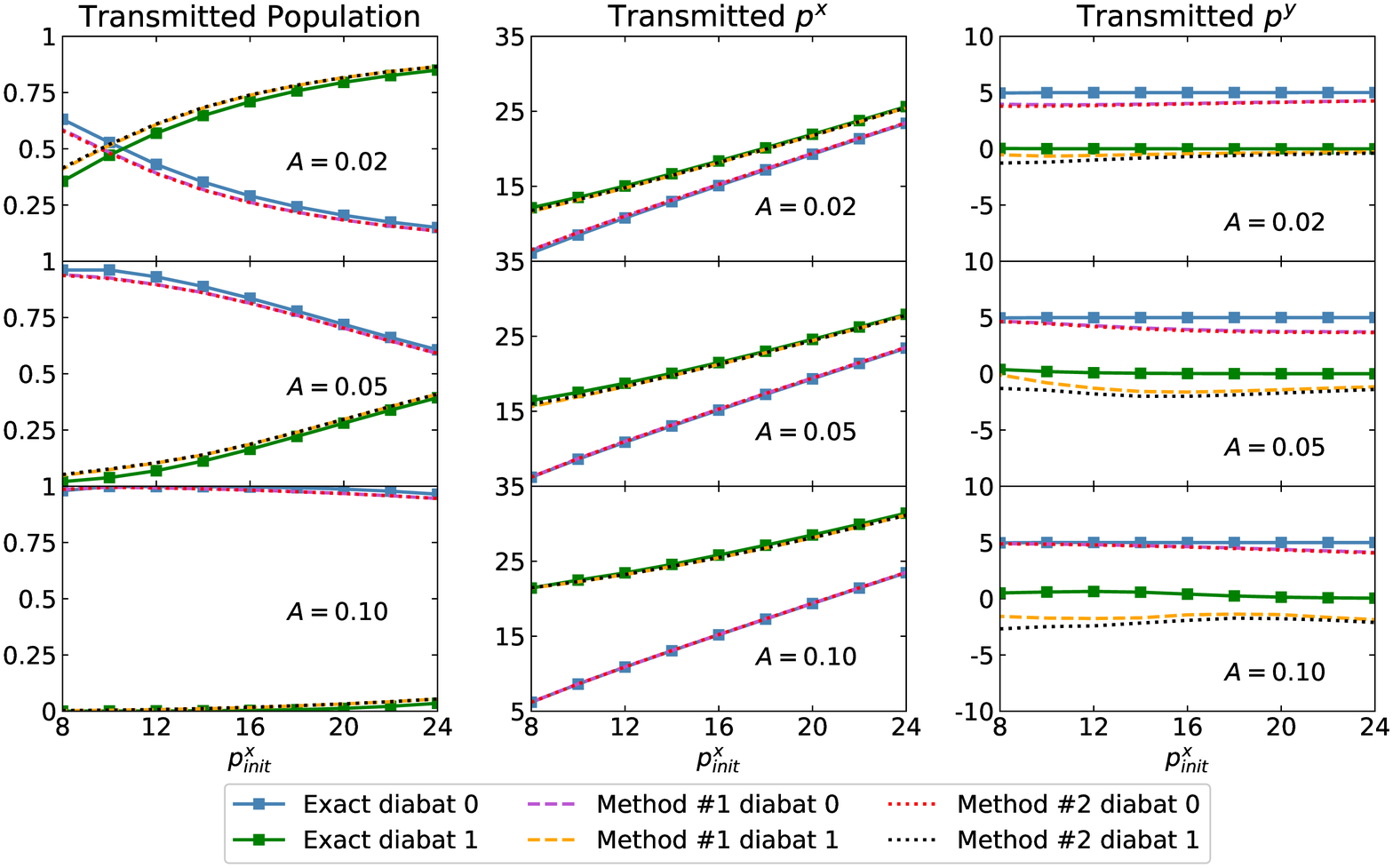}
    	\caption{
            Same as \reffig{fig: smallW_ang0_trans}, but now with $W = 5.0$.
            Left: Transmitted population distribution on the diabatic surfaces after scattering as a function of initial momentum along the $x$-direction, $p^x_{init}$.
            Middle: The $x$-direction momentum on the diabatic surfaces as a function of $p^x_{init}$.
            Right: The $y$-direction momentum on the diabatic surfaces as a function of  $p^x_{init}$. 
            The system is initialized with $\bm{p}_{init} = (p^x_{init}, 0)$. 
            FSSH predicts the correct population as well as the $x$-direction momentum for each diabatic surface, but FSSH is only partially correct for the $y$-direction momentum. 
            The difference between the two rescaling {\em ans\"{a}tze} is negligible for this case. 
            }
    	\label{fig: mediumW_ang0_trans}
    \end{figure}

    Next, in \reffig{fig: mediumW_ang0_trans}, we investigate the same case but with a larger $W$: We set $W = 5.0$ and plot the same observables as in \reffig{fig: smallW_ang0_trans}. 
    As far as the accuracy of FSSH is concerned, our conclusions are the same as for the case of $W = 0.5$.
    Both FSSH {\em ans\"{a}tze} capture the correct $p^x$ and the approximately correct $p^y$. The error increases as $A$ increases. 
    For this model problem, the exact momentum change in the $y$-direction (for wave packets that do not switch adiabats) is again equal to $W$, only now $W = 5.0$.
    \par
\begin{figure}[h!]
    	\centering
    	\includegraphics[width=6in]{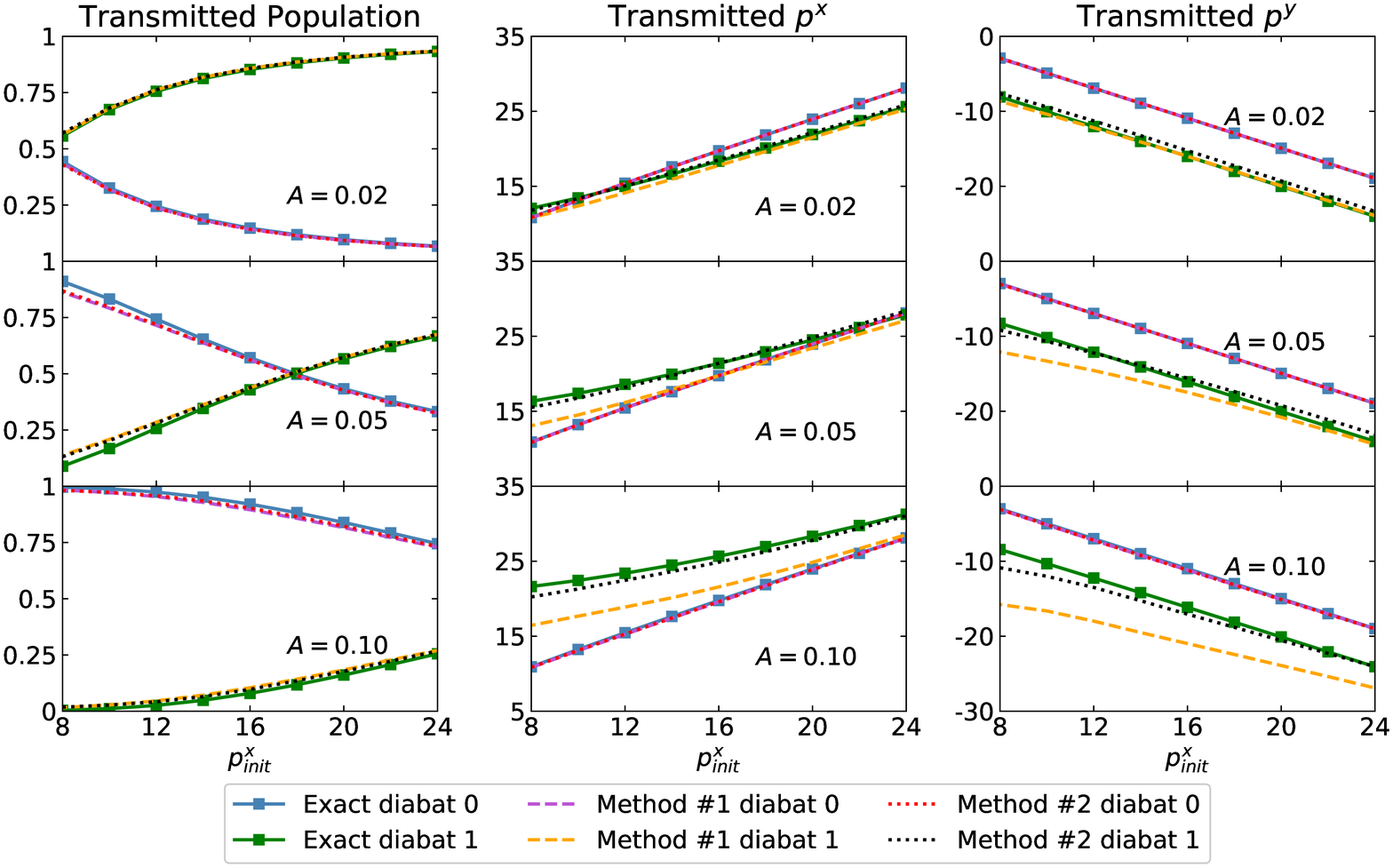}
    	\caption{
            Same as \reffig{fig: smallW_ang0_trans}, but now with $W = 5.0$ and $\bm{p}_{init} = (p^x_{init}, -p^x_{init})$.
            Left: Transmitted population distribution on the diabatic surfaces after scattering as a function of initial momentum along the $x$-direction, $p^x_{init}$.
            Middle: The $x$-direction momentum on the diabatic surfaces as a function of $p^x_{init}$.
            Right: The $y$-direction momentum on the diabatic surfaces as a function of  $p^x_{init}$. 
            The system is initialized with $\bm{p}_{init} = (p^x_{init}, -p^x_{init})$. 
            While both FSSH {\em ans\"{a}tze} predict the correct population, the momentum results of Method \#2 are clearly better than those of Method \#1. }
    	\label{fig: mediumW_ang-1_trans}
    \end{figure}

    Finally, for a meaningful comparison of the two rescaling {\em ans\"{a}tze} and as a means of differentiation, we turn to an alternative set of initial momentum conditions: All trajectories are initialized with a momentum $\bm{p}_{init} = (p^x_{init}, -p^x_{init})$. 
    As plotted in \reffig{fig: mediumW_ang-1_trans}, the results of two {\em ans\"{a}tze} become different: Method \#2 almost captures the correct momentum distribution, while Method \#1 consistently underestimates $p^x$ and overestimates $p^y$ on diabat 1. 
    From this observation, we empirically infer that, if surface hopping is applicable with complex Hamiltonians, Method \#2 must be more physically meaningful than Method \#1.
    Evidently, the optimal rescaling direction is Method \#2, which does not depend on any dynamical information.
    
\section{Discussion } \label{sec: discussion_and_conclusion}

    \subsection{Understanding Berry's Magnetic Force: The case of small or moderately sized $W$ } \label{sec: understanding berry's force}
        To incorporate Berry's phase effects into semiclassical dynamics, we have used the well-known magnetic force {\em ansatz} \cite{shankar_principles_2012} in \refeq{eq: magnetic force}.
        To better understand this force in the context of semiclassical dynamics, note that, in \reffigs{fig: smallW_ang0_trans} and \ref{fig: mediumW_ang0_trans}, one finds that the exact momentum change in the $y$-direction is independent of $A$ and $p^x_{init}$.  
        In fact, as long as $W$ is not too large, if the system is initialized on the upper surface, one always ends up with $p^y_1 = W$ and $p^y_0 = 0$; if the system is initialized on the lower surface, one always ends up with $p^y_1 = 0$ and $p^y_0 = -W$. 
        Neither $A$ or $p^x_{init}$ has an effect on the $p^y$ results.
        These features are completely consistent with the fact that Berry's phase is a topological (rather than dynamic) effect. 
        Nevertheless, in \reffigs{fig: smallW_ang0_trans} and \ref{fig: mediumW_ang0_trans}, one finds that Berry's topological phase has clear dynamic consequences.
        \par
        Within the context of semiclassical dynamics, the {\em ansatz} of an extra magnetic force can partially handle these effects: Given the expression for $\bm{F}^{mag}_1$ in \refeq{eq: Fmag anal}, if a trajectory is initialized on the upper surface and propagated {\em adiabatically, i.e. without any hopping and assuming full transmission}, we find that the final momentum in the $y$-direction is:
			\begin{equation} \label{eq: py1 anal}
			\begin{aligned}
				p^y_1 &= \int_{t=0}^{t=\infty} \frac{\hbar}{2m}\partial_x\theta\partial_y\phi\sin{\theta}p^x dt = \frac{\hbar \partial_y\phi}{2} \left.\cos{\theta}\right|_{\theta(t=0)}^{\theta(t=\infty)} = \hbar\partial_y\phi = \hbar W
			\end{aligned}
			\end{equation}
		Similarly, if a trajectory is initialized on the lower surface and propagated without hopping, we recover
			\begin{equation} \label{eq: py0 anal}
			\begin{aligned}
			p^y_0 &= -\hbar W
			\end{aligned}
			\end{equation}
        Clearly, \refeqs{eq: py1 anal} and \ref{eq: py0 anal} are effectively the correct, semiclassical adiabatic limits; our {\em ansatz} for including Berry's forces within FSSH appears reasonable.
        \par
        Let us next address the question of whether semiclassical FSSH is trustworthy in practice in the limit of {\em finite} (or nonzero) hopping probabilities. In \reffig{fig: smallW_ang0_trans}, in the case of a small $W$ value, we saw that FSSH almost captures the correct results but the agreement is not perfect. 
        As we will show now, this non-agreement can be traced back to the very basic concept of independent FSSH trajectories with variable hopping positions.
        Consider for a moment the early surface hopping proposal by Tully and Preston \cite{tully_trajectory_1971}, whereby a trajectory hops between adiabats only at a crossing point in the spirit of Laudau-Zener transition. 
        In this case, it is quite easy to see that surface hopping should be nearly exact. 
        On the one hand, for a trajectory that does not hop at the crossing point, the ending $p^y$ will be exactly $W$ given the limit of zero hopping (see \refeq{eq: py1 anal}). 
        On the other hand, for a trajectory that hops at exactly the crossing point, half of the transmitted trajectory will run on one adiabat and half will run on the other adiabat. Thus, by symmetry of the Berry's force, {\em i.e.}, the fact that  $\bm{F}^{mag}_0 = -\bm{F}^{mag}_1$, the final $p^y$ will be $0$. 
        Therefore, Tully-Preston surface hopping must be accurate for incorporating geometric phase, and any deviations in the FSSH the $p^y$ results must be caused by the fact that Tully's FSSH algorithm allows trajectories to hop up and down, back and forth, multiple times in the coupling region; 
        this complicated hopping picture no longer guarantees that the $y$-momentum induced by the Berry magnetic force will be accurate.
        In the end, the small inaccuracies in \reffigs{fig: smallW_ang0_trans} and \ref{fig: mediumW_ang0_trans} appear inevitable if one sticks with the independent FSSH algorithm, even in the limit of small $W$.

        \subsection{The Limitations of the Modified FSSH} \label{sec: The limitations of the modified FSSH}
        Next, let us consider larger $W$ values and/or non-perpendicular incoming velocities (so that $\bm{F}_1^{mag}$ is negative in the $x$-direction), where another feature can also appear: Reflection. 
        Even though the adiabats are entirely flat, it is possible to observe reflection!
			\begin{figure}[h!]
				\centering
				\includegraphics[width=6in]{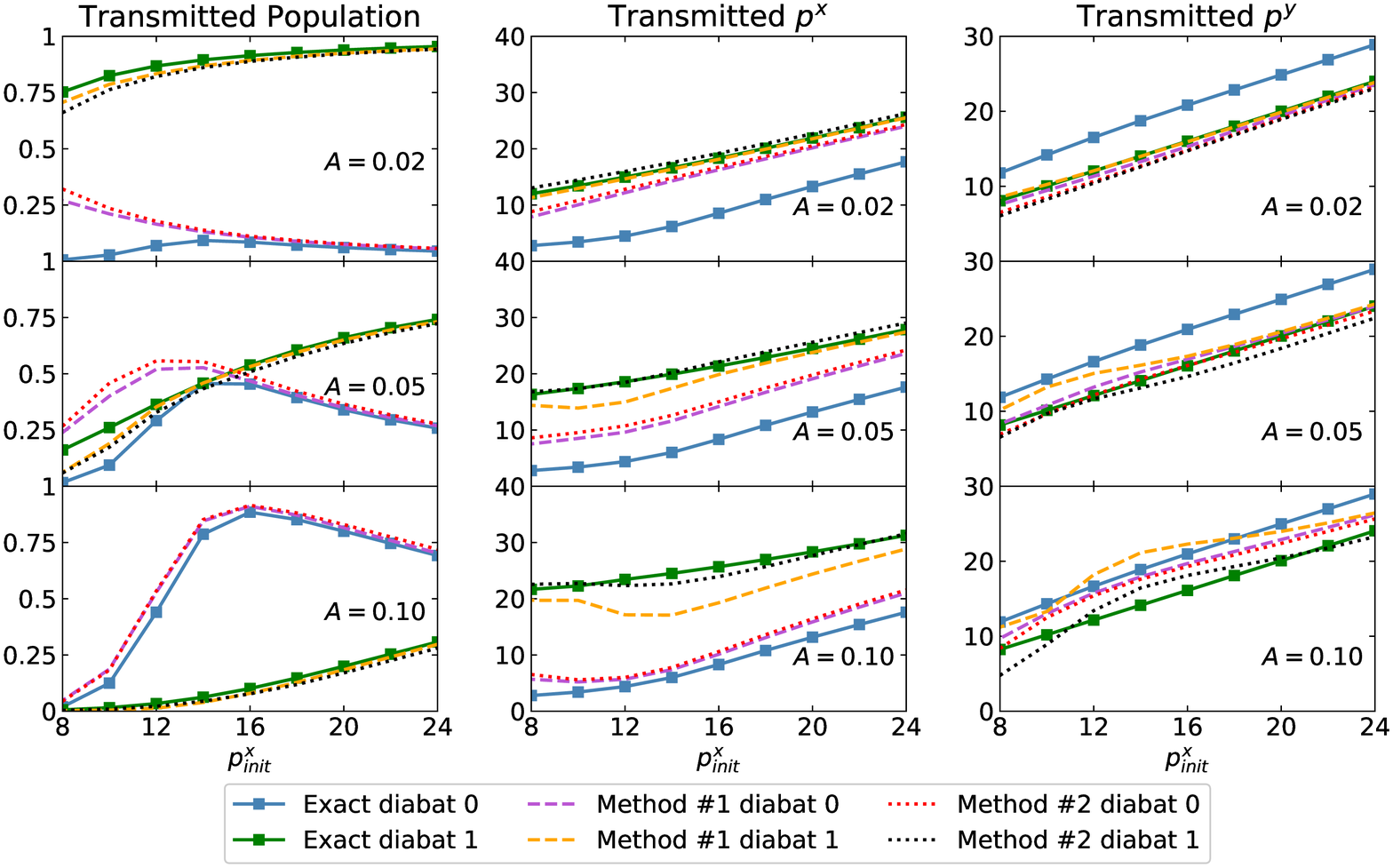}
				\caption{
                    Same as \reffig{fig: smallW_ang0_trans}, but now with $W = 5.0$ and $\bm{p}_{init} = (p^x_{init}, p^x_{init})$.
                    Left: Transmitted population distribution on the diabatic surfaces after scattering as a function of initial momentum along the $x$-direction, $p^x_{init}$.
                    Middle: The $x$-direction momentum for transmission part on the diabatic surfaces as a function of $p^x_{init}$.
                    Right: The $y$-direction momentum for the wave packets transmitted on the diabatic surfaces as a function of  $p^x_{init}$.
                    The system is initialized with $\bm{p}_{init} = (p^x_{init}, p^x_{init})$. 
                    In this case, FSSH with Method \#2 is still better than Method \#1. 
                    When $p_x^{init}$ is small, no FSSH results are accurate.
                }
				\label{fig: mediumW_ang1_trans}
			\end{figure}
			\begin{figure}[h!]
				\centering
				\includegraphics[width=6in]{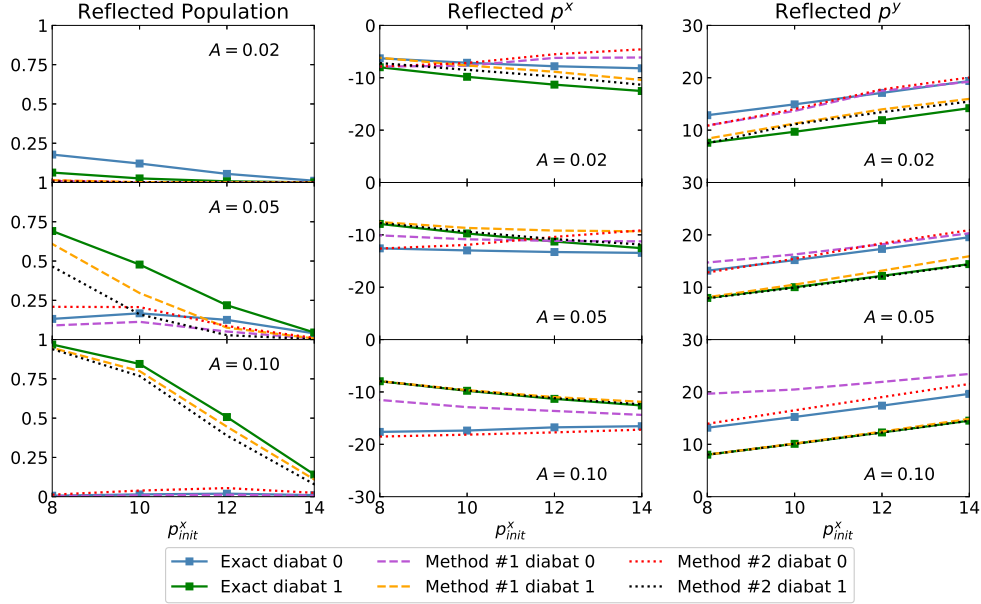}
				\caption{
                    Same as \reffig{fig: mediumW_ang1_trans}, but now the reflected components are plotted.
                    Left: Reflected population distribution on the diabatic surfaces after scattering as a function of the initial momentum along the $x$-direction, $p^x_{init}$.
                    Middle: The $x$-direction momentum for the reflected component on the diabatic surfaces as a function of $p^x_{init}$.
					Right: The $y$-direction momentum  for the reflected component on the diabatic surfaces as a function of  $p^x_{init}$. 
					The system is initialized with $\bm{p}_{init} = (p^x_{init}, p^x_{init})$. 
                    Both exact dynamics and FSSH predict reflection when $p^x_{init}$ is small.
                }
				\label{fig: mediumW_ang1_refl}
			\end{figure}
        In \reffigs{fig: mediumW_ang1_trans} and \ref{fig: mediumW_ang1_refl}, we let $p^y_{init} = p^{x}_{init}$ and we investigate both the transmitted and reflected particles, respectively. 
        We find that both the exact quantum solution and the modified FSSH algorithms predict some amount of reflection provided that we apply the correct magnetic force in our FSSH algorithm. 
		That being said, although Method \#2 is still better than Method \#1, neither method can fully capture the correct population and momentum quantitatively even when $A$ is small. 

        Finally, let us address the case of very large $W$.
		\begin{figure}[h!]
			\centering
			\includegraphics[width=6in]{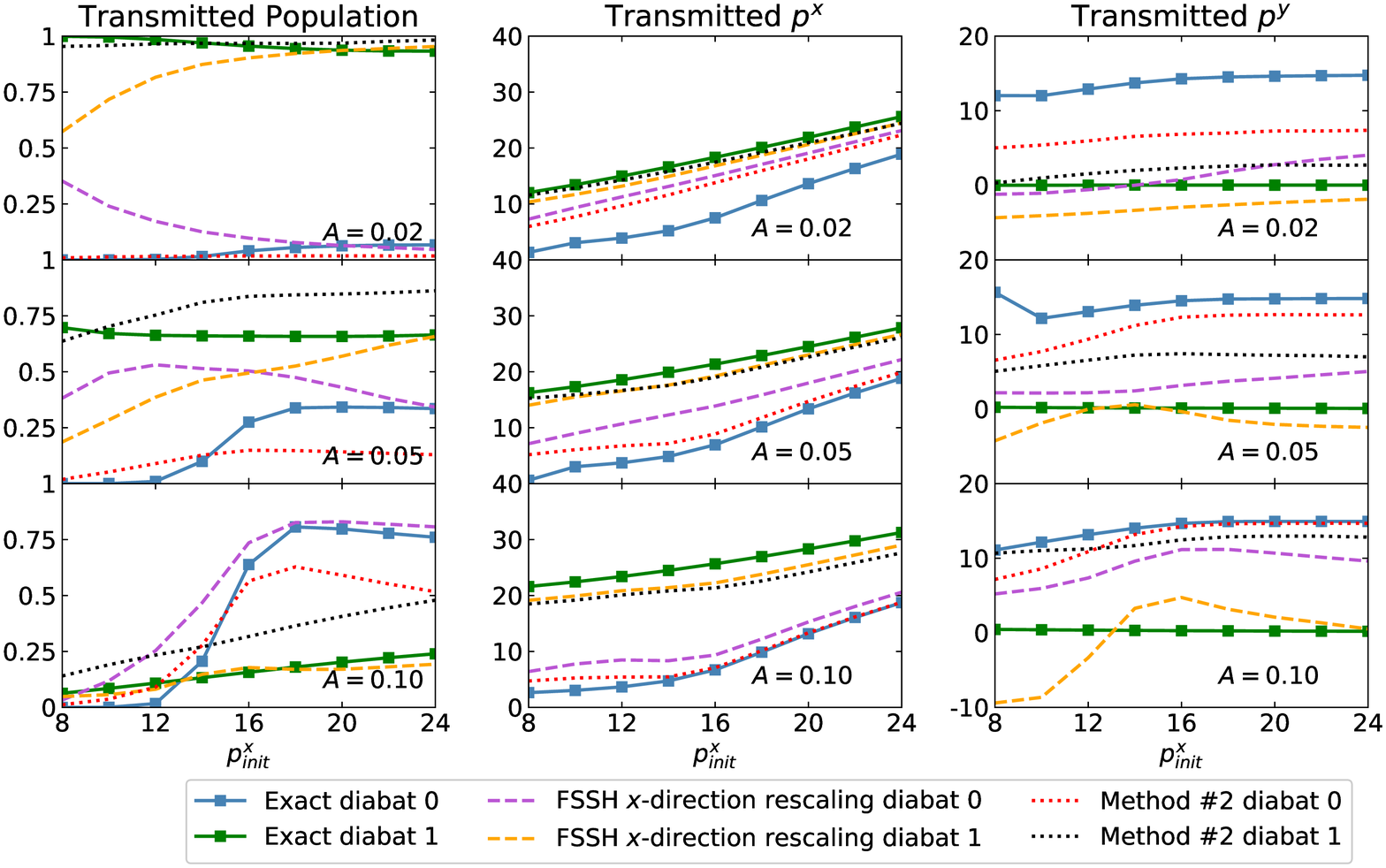}
			\caption{
                Same as \reffig{fig: smallW_ang0_trans}, but now with $W = 15.0$.
                Left: Transmitted population distribution on the diabatic surfaces after scattering as a function of initial momentum along the $x$-direction, $p^x_{init}$.
                Middle: The $x$-direction momentum on the diabatic surfaces as a function of $p^x_{init}$.
                Right: The $y$-direction momentum on the diabatic surfaces as a function of  $p^x_{init}$. 
                The system is initialized with $\bm{p}_{init} = (p^x_{init}, 0)$. 
                For the large value of $W$ studied here, neither rescaling in the $x$-direction nor Method \#2 is quantitatively accurate.
            }
			\label{fig: largeW_ang0_trans}
		\end{figure}
		In \reffig{fig: largeW_ang0_trans}, we plot simulation results for $W = 15$. 
        For this case, an important nuance arises regarding to our FSSH algorithm. 
        Unlike the case of small or medium $W$, where Method \#2 is equivalent to rescaling in the $x$-direction, for the case of large $W$, Method \#2 can actually rescale momenta along the $x$-direction for some coordinates but along the $y$-direction for others (see \refeq{eq: method 2 in model}).
        As a means of assessing this unusual {\em ansatz}, we will introduce yet another rescaling scheme: Simple rescaling along the $x$-direction after a surface hop. 
        \par
        From the results in \reffig{fig: largeW_ang0_trans}, we find that, when $W$ is large, no modified FSSH algorithm works well.\footnote{ Method \#1 also fails (not shown). }
        One is not even able to capture the electronic state populations as a function of $p^x_{init}$.
        One can conceive of two possible explanations for this dramatic failure: 
        $(i)$ When $W$ is large, the complex Hamiltonian matrix oscillates rapidly with frequency $W$ as a function of the coordinate $y$, and so the dynamics may be outside the classical region, and quantum effects may be essential, as in the case of a time-dependent Hamiltonian with large frequency $\omega$. 
        $(ii)$ It is also possible that we have not yet found the optimal approach for velocity rescaling after a hop.
        Understanding how and why FSSH fails in the case of large $W$ deserves further investigation.

        \subsection{ Time Dynamics }\label{sec: time dynamics}
        Before concluding, let us turn to time dynamics rather than scattering probabilities. 
        So far in this manuscript, we have focused on the asymptotic states {\em after} a scattering event -- rather than the time dynamics of the underlying wave function {\em during} the scattering event.
		\begin{figure}[h!]
			\centering
			\includegraphics[width=6in]{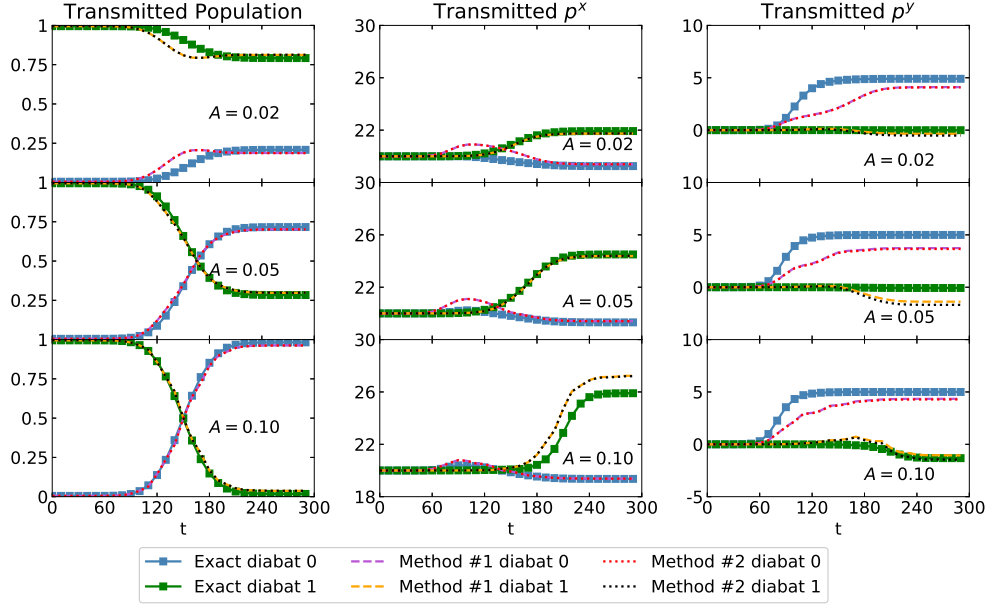}
			\caption{
                Time evolution of population and momenta with $W = 5.0$.
				Left: Electronic population on the diabatic surfaces as a function of time $t$.
				Middle: The $x$-direction momentum on the diabatic surfaces as a function of time.
				Right: The $y$-direction momentum on the diabatic surfaces as a function of time.
				The system is initialized with $\bm{p}_{init} = (20, 0)$.
                The initial electronic population is chosen to be $n_0 = 0.005$ and $n_1 = 0.995$ (rather than $n_1 = 1$) in order to avoid intense oscillations when converting from the adiabatic to the diabatic representation. 
				For the population dynamics, FSSH results are quite accurate.
                With regards to momentum, FSSH is reasonably accurate.
			}
			\label{fig: dyn_diab}
		\end{figure}
        To better understand the dynamics, in \reffig{fig: dyn_diab} we plot the time evolution for the populations and momenta on the diabats. The initial momentum $\bm{p}_{init}$ is set to be $(20,0)$. To obtain FSSH statistics on the diabatic surfaces, we use method 3 from Ref. \onlinecite{landry_communication:_2013}. 
        This conversion method leads to intense oscillations if the wavefunction is initialized entirely on the upper diabatic surface, and to avoid such a numerical issue, we initialize the wavefunction with a slight superposition state: $99.5\%$ of the population is initialized on the upper diabatic surface, while $0.5\%$ are initialized on the lower one.
        From the data in \reffig{fig: dyn_diab}, we find that, despite the fact that the scattering process is dynamically complicated, FSSH dynamics are not actually that bad (just as for real Hamiltonians): The population dynamics predicted by FSSH are reasonably accurate, and the overall trend of momentum dynamics are basically in agreement with the exact dynamics.
		\begin{figure}[h!]
			\centering
			\includegraphics[width=6in]{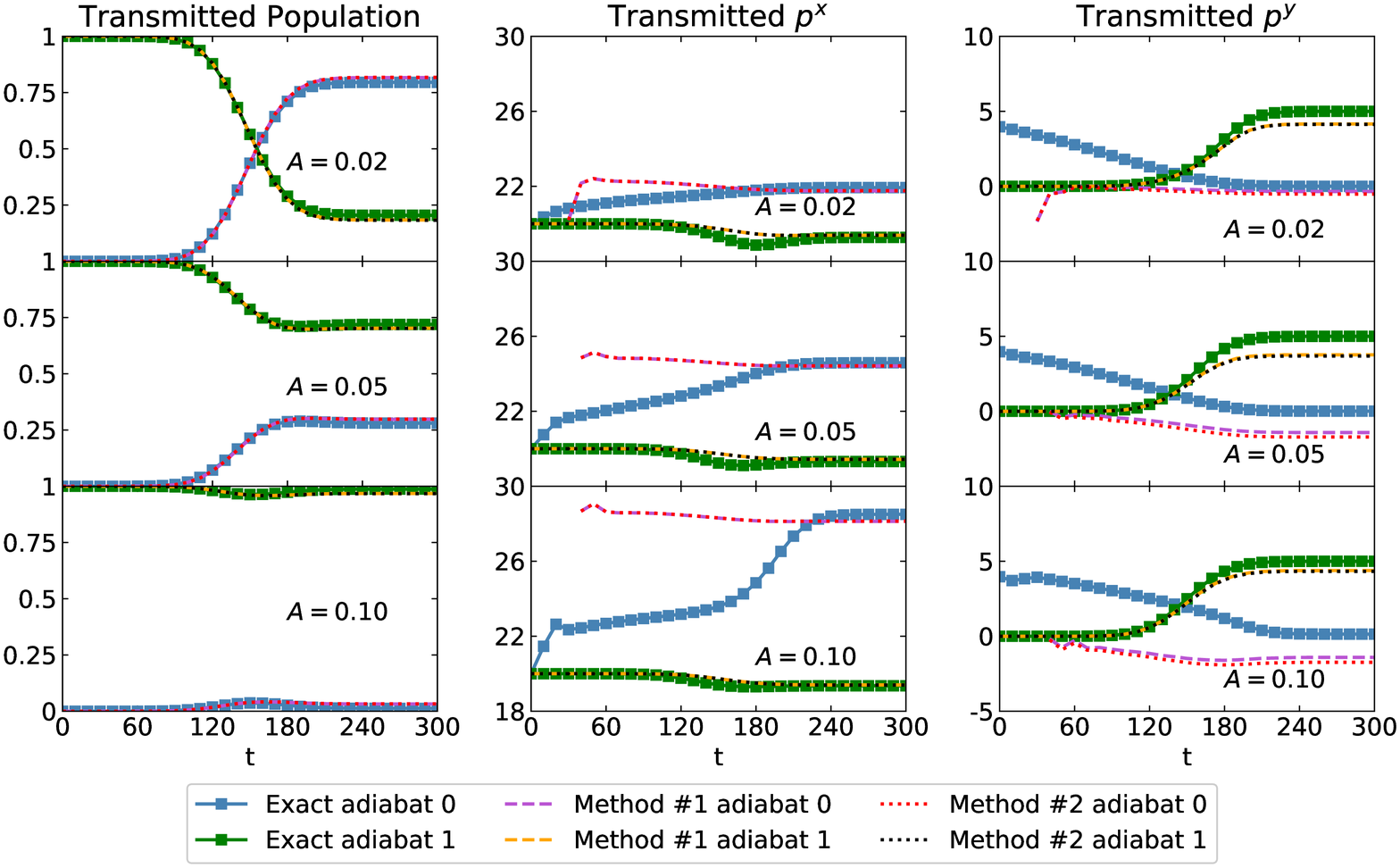}
			\caption{
                $W = 5.0$. 
                Left: Electronic population on the adiabatic surfaces as a function of time $t$.
				Middle: The $x$-direction momentum on the adiabatic surfaces as a function of time.
				Right: The $y$-direction momentum on the adiabatic surfaces as a function of time.
				The system is initialized with $\bm{p}_{init} = (20, 0)$. 
				Some early FSSH momentum for the lower adiabatic surface is missing because no trajectory is on that surface. 
				Although FSSH captures the correct time evolution of population as well as approximately correct ending momentum, it fails to predict the correct momentum as a function of time. 
            }
			\label{fig: dyn_adiab}
		\end{figure}
        \par
        Lastly, let us turn to the adiabatic representation. 
        To generate exact adiabatic momenta, we rotate the electronic wavefunctions from the diabatic representation to the adiabatic representation, using the analytical eigenvectors in \refeq{eq: eva evt dc anal}. 
        Note that the quantities $\matrixel{\psi_0}{\bm{\nabla}}{\psi_0}$ and $\matrixel{\psi_1}{\bm{\nabla}}{\psi_1}$ are usually non-zero (and of course purely imaginary).
        The contribution of these terms must be included when evaluating the momentum on the adiabatic surfaces.
        Thus, if the exact wavefunction is $\ket{\Psi} = C_0\ket{\psi_0} + C_1\ket{\psi_1}$, we estimate the exact momentum on the upper adiabatic surface to be
        \begin{equation} \label{eq: upper adiab momentum}
        \begin{aligned}
            \bm{p}_1
            &= \frac{\matrixel{C_1\psi_1}{-i\hbar\bm{\nabla}}{C_1\psi_1}}{\bracket{C_1\psi_1}{C_1\psi_1}} \\
            &= -i\hbar (\frac{C_1^*\bm{\nabla}C_1}{C_1^*C_1} + \matrixel{\psi_1}{\bm{\nabla}}{\psi_1})
        \end{aligned}
        \end{equation}
        \par
        Next, let us consider FSSH. 
        Normally, because FSSH is defined in the adiabatic basis, one would expect FSSH to be most accurate in this representation.
        For FSSH, the adiabatic momenta are computed simply by averaging the momentum of all trajectories on a given adiabatic surface.
        \par
        The dynamics on the adiabatic surfaces are plotted in \reffig{fig: dyn_adiab}. 
        For the adiabatic populations, FSSH again captures accurate dynamics. 
        For the momentum, however, the FSSH result on adiabat 0 is extremely inaccurate for early times. 
        In theory, this error could arise because, at early times, the details of one wave packet spreading from one adiabat to another must reflect the quantum nature of matter waves.
        A simpler and more likely explanation, however, is that FSSH fails here simply because semiclassical dynamics treat $\bm{p}$ classically whereas exact quantum dynamics interprets momentum as a phase change (that can more naturally account for the presence of geometric phase). 
        Either way, it is quite surprising that, physical observables (as calculated by FSSH) in a diabatic basis appear more accurate than those in an adiabatic basis.

	\section{Summary} \label{sec: summary}
		To summarize, we have proposed a modified version of FSSH to incorporate non-adiabatic semiclassical systems with complex Hamiltonians. 
        For a chemistry audience accustomed to non-adiabatic transitions, we have shown how to include complex Hamiltonians and Berry's forces; for a physics audience accustomed to adiabatic dynamics with complex Hamiltonians, we have shown one means to take the non-adiabatic limit and including hopping.
		For motion along adiabatic surfaces, we invoke the usual concept of adiabatic ``magnetic forces" to account for Berry's phase (\refeq{eq: magnetic force expr2}), and some evidence has been provided that this approach is compatible with standard FSSH.
		\footnote{
		Interestingly, we note that, even though FSSH is grounded in the notion of dynamics along adiabats, a few researchers have designed surface hopping schemes in a diabatic basis.\cite{hack_semiclassical_2000,wang_fewest_2015}
        Within such a diabatic framework, one would {\em not} be able to use ``adiabatic magnetic forces" to account for Berry's phase (as we have done here). 
        Instead, one would need to account for Berry's phase when adjusting velocities after a hop, and there is no guarantee that such an approach would be robust. 
		}$^,$\cite{hack_semiclassical_2000,wang_fewest_2015}
        For the momentum rescaling scheme, we compare three potential {\em ans\"{a}tze} and show that Method \#2 is the best rescaling scheme:
        after a hop $j \rightarrow k$, the momentum should be adjusted in the direction $Re(e^{i\eta} \bm{d}_{jk})$ where $\eta$ is chosen to maximize $||Re(e^{i\eta} \bm{d}_{jk})||^2$, which is the same direction as $\bm{\nabla}\theta$ for a two-state model.
        \footnote{
            Note that, if $W$ is large enough, as discussed in the context of \reffig{fig: largeW_ang0_trans}, Method \#2 and $\bm{\nabla}\theta$ may give different rescaling directions. Nevertheless, for such large $W$ values, FSSH does not appear to be accurate -- again, see \reffig{fig: largeW_ang0_trans} -- and so Method \#2 would appear to be a robust {\em ansatz} that should be applicable for {\em ab initio} calculations.
        }
        Evidently, choosing a dynamical rescaling direction is not appropriate.
        \par
        With these adjustments, our overall conclusion is that, a modified FSSH algorithm can capture many important non-adiabatic dynamical features ({\em e.g.} the scattering probabilities and the approximate scattering momenta), but FSSH cannot capture a few features ({\em e.g.} the detailed early time dynamics of momentum transfer).

	\section{Open Questions} \label{sec: open questions}
        With the above summary in mind, several questions now present themselves. 
        On the practical side, the first methodological question one must pose is: 
        Have we constructed the optimal FSSH algorithm or is there another, better option available for the case of a complex electronic Hamiltonian? 
        Considering the errors in the $y$-momentum in \reffigs{fig: smallW_ang0_trans} and \ref{fig: mediumW_ang0_trans} and the discussion of independent trajectories in \refsec{sec: understanding berry's force}, we note that Truhlar {\em et al} have constructed an FSSH algorithm with time uncertainty\cite{jasper_fewest-switches_2002} which was designed to introduce a small amount of time non-locality.
        Would a similar approach help improve FSSH in this case and reduce the number of hops in the coupling region? 
        Or is it simply impossible to model Berry's phase well with independent trajectories, given that Berry's phase is geometric and topological (and therefore intrinsically non-local)?
        \par
        Second, again on the practical side, a modern FSSH implementation can avoid calculating derivative couplings unless a hop is required;\cite{jain_efficient_2016}
        as far as propagating time dependent Sch\"{o}dinger equation, $\bm{d}\cdot\bm{p}/m$ is enough. Unfortunately, in the case of a complex Hamiltonian with Berry's forces, apparently one must calculate $\bm{d}$ at every time step in order to evaluate $\bm{F}^{mag}_j$.
        One must wonder: Is there a practical and efficient approach to construct such a Berry force easily, ideally a scheme that will be stable with a large number of electronic states and will avoid the trivial crossing problem?\cite{fernandez-alberti_identification_2012,nelson_artifacts_2013,plasser_surface_2012,wang_simple_2014,meek_evaluation_2014,jain_efficient_2016}
        \par
		Third, on the theory side, one must also wonder: Can any of our proposed extensions of FSSH be tied back to a more rigorous theory of quantum mechanics?  For the case of a real electronic Hamiltonian, our research group and the Kapral research group have successfully tied FSSH back to the QCLE\cite{kapral_surface_2016,subotnik_can_2013}. However, the QCLE is a first order expansion that cuts off at zeroth order in $\hbar$, whereas Berry's phase requires a second-order expansion: Note that the magnetic force in \refeq{eq: magnetic force expr2} is first order in $\hbar$. Can we relate an extended version of FSSH to an extended version of the QCLE for the case of complex Hamiltonians?
        \par
        Fourth, according to \reffigs{fig: mediumW_ang1_trans} and \ref{fig: mediumW_ang1_refl}, the magnetic force in \refeq{eq: magnetic force expr2} can lead to wave packet separating as trajectories on different adiabatic surfaces are turned in different directions, some transmitted and some reflected. 
        Thus, the sharp reader will no doubt isolate yet another question.
        Recall that, when deriving FSSH from the QCLE, the question of decoherence and wave packet separation arises naturally.\cite{kapral_surface_2016} After all, wave packets on different adiabatic surfaces feel different static, adiabatic forces that lead to separation eventually; 
        and for years, many researchers have constructed practical solutions for incorporating decoherence into FSSH to account for such effects.\cite{bittner_quantum_1995,schwartz_quantum_1996,fang_improvement_1999,subotnik_decoherence_2011,wang_recent_2016,hack_electronically_2001,volobuev_continuous_2000,jasper_electronic_2005,prezhdo_mean-field_1997}
        For the present paper, however, we now see a new phenomenon: With Berry's forces, wave packet separation is caused by wave packets on different surfaces feeling different magnetic forces that depend on velocity.
        Furthermore, these ``magnetic" forces appear only in the strong coupling region, which negates our usual understanding of decoherence being a phenomenon that emerges after wave packets pass through coupling region and only thereafter move apart in different directions. \cite{zhu_coherent_2004}
        Thus, another immediate question is how should we appropriately model such magnetically induced decoherence within FSSH so as to recover the correct dynamics.
        \par
        Given the inherent difficulties of including decoherence within FSSH, the questions above lead to a fifth question: Is it possible that a different mixed quantum classical scheme might strongly outperform FSSH for the case of complex Hamiltonians? 
        In particular, for problems of decoherence, {\em ab initio} multiple spawning (AIMS) is a more natural {\em ansatz}.\cite{martinez_multi-electronic-state_1996,ben-nun_ab_2000}
        And yet, AIMS is most efficient in an adiabatic basis, where single valued wave functions can be difficult to find.
        Interestingly, there has been a great deal of work investigating conical intersection's geometric phase and choice of basis within AIMS for real Hamiltonians, and the overall conclusion appear to be that we should run dynamics with electronic wavefunctions chosen at a single location.\cite{meek_best_2016}
        Thus, one can ask, can the results in Ref. \onlinecite{meek_best_2016} for adiabatic AIMS be easily extended to work with complex Hamiltonians?
        \par
        The final, sixth question is perhaps most exciting of all.
        On the experimental front, one must wonder: Can any of the dynamics predicted in \refsec{sec: results} above be detected experimentally? 
        For instance, the numerical model above suggests that, whenever an electronic transition (in the $x$-direction) occurs between two electronic states with spin, one ought to find a signature of nuclear or vibrational motion (in the $y$-direction) as arising from Berry's phase for the case of a molecule in a magnetic field -- provided that the transition occurs in the normal regime where an electron changes character along a single adiabat. 
        {\em Vice versa}, no such signature should be observable for a transition in the inverted regime where an electron changes character but the adiabat also changes.  Can this dichotomy be seen experimentally?
        Can we find realistic molecular systems with large enough susceptibilities such that, in very large magnetic fields, we will observe dynamical Berry phase effects?  
        Or, if we recall that Marcus theory assumes a threshold amount of nuclear friction, a pessimist must ask: Will the inevitable presence of some nuclear friction eliminate all such effects?
        And lastly, how will these features behave when the complex phase is more complicated, so that $\partial_y\phi$ is not a constant (as assumed above)?
        These fascinating experimental and theoretical questions will hopefully be answered in the near future.
		

\appendix
\section{Method \#3 Results} \label{apdx: method 3 results}
    Here we briefly present scattering results for Method \#3 in \ref{sec: direction of momentum rescaling}. 
     \begin{figure}[h!]
        \centering
        \includegraphics[width=6in]{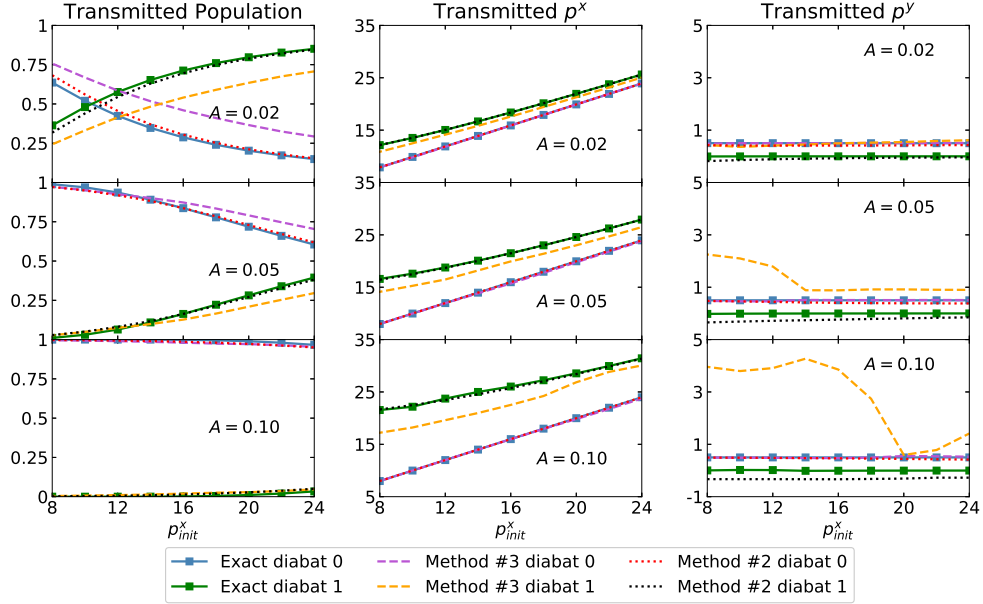}
        \caption{
            Same as \reffig{fig: smallW_ang0_trans}, but now we compare Methods \#2 and \#3.
            Note that Method \#3 results are far worse than results using Method \#2.
        }
        \label{fig: smallW_ang0_trans_rescalecd}
    \end{figure}
    In \reffig{fig: smallW_ang0_trans_rescalecd}, we report results for the case $W = 0.5$ with Methods \#3 and \#2. 
    Clearly, the results indicate that neither the correct population nor the correct momentum can be captured by Method \#3.
    Now, at first glance, it might seem that the $y$-momentum on diabat 0 is correctly captured. 
    This instinct is merely an illusion, however, as this ``accurate" results is caused only surreptitiously from the fact that, at the end of the simulation, the particles remaining on the upper adiabatic surface ({\em i.e.} diabat 0) are mostly those trajectories that never hop, and so the average $y$-momentum will always go to the correct answer (as induced by the magnetic force). 
    This correct answer is the zero hopping limit (or adiabatic limit) as discussed in \refsec{sec: understanding berry's force} that arises from simple classical mechanics (and ignoring all surface hops). 
    \par
    Overall, even though it might appear natural, Method \#3 does not coincide with the correct physical picture.

\section{Method \#2 with a General Derivative Coupling} \label{apdx: method 2 with a general derivative coupling}
    Here, we analyze Method \#2 for a general, two-state diabatic problem.
    We denote a general (complex) derivative coupling vector as $\bm{d} \equiv \bm{d}_{R} + i\bm{d}_{I}$.
    When using Method \#2, we maximize the following within the interval $\eta \in [0, \pi)$:
    \begin{equation}\label{eq: method 2 quantity to maximize}
    \begin{aligned}
        f(\eta) &= ||Re(e^{i\eta} \bm{d})||^2 = ||\cos{\eta}\bm{d}_{R} - \sin{\eta}\bm{d}_{I}||^2 \\
        &= \frac{1}{2}\left(||\bm{d}_{R}||^2 + ||\bm{d}_{I}||^2\right) + \frac{\cos{2\eta}}{2}\left(||\bm{d}_{R}||^2 - ||\bm{d}_{I}||^2\right) - \sin{2\eta}\bm{d}_{R} \cdot \bm{d}_{I} \\
    \end{aligned}
    \end{equation}
    Setting $f'(\eta) = 0$ tells us $\eta$ should satisfy 
    \begin{equation}\label{eq: method 2 first derivative}
    \begin{aligned}
        \tan{2\eta} = \frac{-2 \bm{d}_{R} \cdot \bm{d}_{I} }{||\bm{d}_{R}||^2 - ||\bm{d}_{I}||^2}
    \end{aligned}
    \end{equation}
    Within the $[0,\pi)$ interval, there exist two solutions: $\eta_0$ and $\eta_1 = \eta_0 + \pi/2$, with $\eta_0 \in [0, \pi/2)$. 
    Using the second derivative $f''(\eta) < 0$, we find that maximizing $f(\eta)$ requires that $\eta$ must satisfy
    \begin{equation}\label{eq: method 2 second derivative}
    \begin{aligned}
        \cos{2\eta} \left(||\bm{d}_{R}||^2 - ||\bm{d}_{I}||^2\right) > 0
    \end{aligned}
    \end{equation}
    Thus, for a general $\bm{d}$, the solution for $\eta$ should satisfy both \refeq{eq: method 2 first derivative} and \refeq{eq: method 2 second derivative}.

\newpage

\bibliographystyle{apsrev}

\end{document}